\documentclass[aps,twocolumn,english,superscriptaddress,citeautoscript,showkeys,preprintnumbers,amsmath,amssymb,floatfix,footinbib,prb]{revtex4-2}
\usepackage{graphicx}
\usepackage[dvipsnames]{xcolor}
\usepackage{soul}
\usepackage{amssymb,amsmath}
\usepackage{hyperref}
\hypersetup{colorlinks=true,linkcolor=red,citecolor=blue}
\usepackage{cleveref}
\usepackage[utf8]{inputenc}
\usepackage[english]{babel}
\usepackage{txfonts}
\usepackage{booktabs}

\DeclareUnicodeCharacter{0303}{HERE!HERE!}
\usepackage{multirow}
\usepackage{listings}
\lstdefinelanguage{Julia}{
  morekeywords={using, module, import, export, function, end, if, else, elseif, while, for, break, continue, return, try, catch, finally, struct, mutable, macro, quote},
  sensitive=true,
  morecomment=[l]{\#},
  morestring=[b]"
}

\lstdefinestyle{mystyle}{
    backgroundcolor=\color{gray!10},
    basicstyle=\scriptsize\ttfamily,
    numbers=left,
    numbersep=0.5em,
    xleftmargin=1em,
    keepspaces=true,
    columns=flexible
}
\usepackage{subcaption}

\usepackage{placeins}

\usepackage{xr}

\newcommand{\IsoME}{\textsc{IsoME}}

\newcommand{\Tc}{$T_\text{c}$}
\newcommand{\TcAD}{$T_\text{c}^\text{AD}$} 
\newcommand{\TcML}{$T_\text{c}^\text{ML}$}

\newcommand{\TcEcdosmu}{$T_\text{c}^{\text{E, cdos, }\mu}$}
\newcommand{\TcEvdosW}{$T_\text{c}^\text{E, vdos, W}$}
\newcommand{\TcEvdosmu}{$T_\text{c}^{\text{E, vdos, } \mu}$}
\newcommand{\Tcexp}{$T_\text{c}^\text{exp}$}

\newcommand{\mustE}{$\mu^{*}_\text{E}$}
\newcommand{\mustAD}{$\mu^{*}_\text{AD}$}

\newcommand{\labh}{LaBH$_8$}

\newcommand{\hs}{H$_3$S}

\newcommand{\wj}{\omega_j}
\newcommand{\wjp}{\omega_{j'}}
\newcommand{\wc}{\omega_\text{c}}

\newcommand{\bbk}{\mathbf k}

\newcommand{\bbkk}{\mathbf {k'}}

\newcommand{\nk}{n\mathbf{k}}
\newcommand{\npkp}{n'\mathbf{k'}}

\newcommand{\afkko}{\alpha^{2} F_{n\mathbf k,n'\mathbf {k'}}(\omega)}

\newcommand{\afo}{\alpha^{2} F(\omega)}

\newcommand{\ef}{\varepsilon_{\mathrm{F}}}
\newcommand{\NF}{N(\ef)}
\renewcommand{\cp}{\mu_\text{F}}     
\newcommand{\nf}{n_\text{F}}    

\newcommand{\Wnkwj}{W_{\nk,\npkp}(i\wj-i\wjp)}
\newcommand{\Wnk}{W_{\nk,\npkp}}

\newcommand{\kB}{k_\text{B}}

\begin{document}

\title{\IsoME{}: 
Streamlining High-Precision Eliashberg Calculations}

\author{Eva Kogler}
\affiliation{Institute of Theoretical and Computational Physics, Graz University of Technology, NAWI Graz, 8010, Graz, Austria}
\author{Dominik Spath}
\affiliation{Institute of Theoretical and Computational Physics, Graz University of Technology, NAWI Graz, 8010, Graz, Austria}
\author{Roman Lucrezi}
\affiliation{Institute of Theoretical and Computational Physics, Graz University of Technology, NAWI Graz, 8010, Graz, Austria}
\affiliation{Department of Materials and Environmental Chemistry, Stockholm University, SE-10691 Stockholm, Sweden}

\author{Hitoshi Mori}
\affiliation{Department of Physics, Applied Physics, and Astronomy, Binghamton University-SUNY, Binghamton, New York 13902, USA}

\author{Zien Zhu}
\affiliation{Mork Family Department of Chemical Engineering and Materials Science, University of Southern California, Los Angeles, California 90089, USA}

\author{Zhenglu Li}
\affiliation{Mork Family Department of Chemical Engineering and Materials Science, University of Southern California, Los Angeles, California 90089, USA}

\author{Elena R. Margine}
\affiliation{Department of Physics, Applied Physics, and Astronomy, Binghamton University-SUNY, Binghamton, New York 13902, USA}

\author{Christoph Heil}
\email[Corresponding author: ]{christoph.heil@tugraz.at}
\affiliation{Institute of Theoretical and Computational Physics, Graz University of Technology, NAWI Graz, 8010, Graz, Austria}

\begin{abstract}
This paper introduces the Julia package \IsoME{}, an easy-to-use yet accurate and robust computational tool designed to calculate superconducting properties. Multiple levels of approximation are supported, ranging from the basic McMillan-Allen-Dynes formula and its machine learning-enhanced variant to Eliashberg theory, including static Coulomb interactions derived from $GW$ calculations, offering a fully \textit{ab initio} approach to determine superconducting properties, such as the critical superconducting temperature (\Tc) and the superconducting gap function 
($\Delta$).  We validate \IsoME{} by benchmarking it against various materials, demonstrating its versatility and performance across different theoretical levels. The findings indicate that the previously held assumption that Eliashberg theory overestimates \Tc~is no longer valid when $\mu^*$ is appropriately adjusted to account for the finite Matsubara frequency cutoff. Furthermore, we conclude that the constant density of states (DOS) approximation remains accurate in most cases. By unifying multiple approximation schemes within a single framework, \IsoME{} combines first-principles precision with computational efficiency, enabling seamless integration into high-throughput workflows through its \Tc{} search mode. This makes \IsoME{} a powerful and reliable tool for advancing superconductivity research. \\

\noindent \textbf{PROGRAM SUMMARY}
\\
\begin{small}
\noindent
{\em Program Title:} \IsoME{}                                          \\
{\em CPC Library link to program files:} (to be added by Technical Editor) \\
{\em Developer's repository link:} https://github.com/cheil/IsoME.jl \\
{\em Licensing provisions:} MIT license (MIT)  \\
{\em Programming language:} Julia 1.10 or higher                                  \\
{\em Supplementary material:}   https://cheil.github.io/IsoME.jl                  \\
{\em Nature of problem:} The challenge addressed by \IsoME{} is the rigorous, first-principles calculation of superconducting properties, particularly the critical temperature (\Tc) and detailed self-energy components. Predicting these properties involves solving the highly nonlinear, coupled Migdal-Eliashberg equations that capture the interplay between electron-phonon interactions and Coulomb repulsion. This is nontrivial because the equations require careful treatment of frequency-dependent interactions, accurate sampling of the electronic density of states near the Fermi level, and efficient summation over extensive Matsubara frequencies. Additionally, incorporating energy-dependent screening effects and ensuring numerical convergence across multiple energy scales further complicates the task. Addressing these issues is essential not only for understanding the fundamental physics of superconductivity but also for guiding the discovery and design of new superconducting materials. \\
{\em Solution method:} \IsoME{} employs isotropic Migdal-Eliashberg theory as its backbone, implementing a hierarchical approach that spans several levels of approximation. At the simplest level, the code uses semi-empirical formulas such as the McMillan-Allen-Dynes equation enhanced by machine learning to provide quick estimates of \Tc. For more detailed studies, it solves self-consistently the full set of Eliashberg equations using either a constant or variable density of states (DOS) approximation, with the possibility to include the full energy-dependent static Coulomb interaction computed via $GW$ methods. The package is written in Julia, ensuring high computational efficiency and ease of integration into high-throughput workflows. It further incorporates sparse-sampling techniques to accelerate Matsubara frequency summations and an automated \Tc{} search mode, thereby balancing computational cost with high-precision predictions.\\
{\em Additional comments including restrictions and unusual features:} \IsoME{} is an open-source Julia package designed for high-throughput superconductivity investigations. Its modular architecture supports diverse approximation schemes that balance efficiency and accuracy. However, robust predictions require high-quality input data and thorough convergence testing, particularly for systems with complex electronic structures or strong energy-dependent interactions near the Fermi level.\\
   \\
\end{small}

\end{abstract}

\date{\today}


\maketitle

\section*{Introduction}
Since the discovery of superconductivity by Onnes in 1911~\cite{onnes_SC_mercury}, unraveling the nature of the superconducting phase has remained a central challenge in condensed matter physics.
A significant milestone was achieved by the Bardeen-Cooper-Schrieffer (BCS) theory ~\cite{BCS-PhysRev.108.1175}, which provided the first microscopic description of weak-coupling superconductors. This concept was later generalized by the Migdal-Eliashberg (ME) theory~\cite{migdal1958,eliashberg1960}, treating the problem in a many-body perturbation approach, to account for effects such as retardation and strong coupling. 
Building on these \textit{ab-initio} methods, simplified semi-empirical equations for the critical temperature (\Tc{}), such as the McMillan~\cite{McMillan2968} and Allen-Dynes~\cite{AllenDynes1975} equations, were formulated as practical approximations. While these equations often provide accurate estimates for \Tc{}, they do not capture the full complexity of the underlying physics. They are limited to predicting \Tc{} alone, neglecting all other superconducting properties relevant for applications, such as single-photon detection~\cite{simon2025}, Josephson junctions \cite{Setzu2008}, and qubits \cite{Nakamura-2011}.

Therefore, until today, ME theory, along with density functional theory for superconductors (SCDFT)~\cite{Oliveira1988SCDFT, Lueders2005_1, Marques2005_2}, remains the state-of-the-art framework for describing superconductors. 

Although ME theory has been available since the 1960s, progress in the field of superconductivity has historically been driven by experimental discoveries. This stems primarily from the computational complexity of the two essential steps required to predict novel superconductors: first, determining the \mbox{(meta-)stable} crystal structure, and second, accurately calculating their superconducting properties.
The exponential growth in computational power over the past decades, however, coupled with the development of highly efficient and accurate numerical codes, has led to a paradigm shift towards the discovery of superconductors \textit{in silico}~\cite{Boeri_JPCM_2021_roadmap}.

The impact of refined computational methods has been strikingly demonstrated with the advent of high-pressure hydride superconductors~\cite{pickars-2020-hydrides,Flores-2020-hydrides,boebinger2024hydride}. The breakthrough began in 2014 with H$_3$S, for which theory predicted a record-high \Tc~\cite{duan_pressure-induced_2014}, soon followed by its independent experimental realization~\cite{drozdov_conventional_2015}. Since then, every hydride superconductor realized experimentally has first been predicted theoretically, underscoring the reliability of computational approaches~\cite{Flores-2020-hydrides,Boeri_JPCM_2021_roadmap}. Notable examples include LaH$_{10}$~\cite{Liu-lah-2017, Somayazulu_lah_2019}, YH$_6$~\cite{Liu-lah-2017, Troyan-yh6-2021}, and LaBeH$_8$~\cite{Song-labh8-2023, Cataldo-labh8-2021, zhang2022design}. These successes stress that carefully conducted crystal structure predictions, combined with accurate \Tc{} computations, serve as trustworthy guides for experimental efforts.

\begin{figure*}[t]
    \centering
    \includegraphics[width=0.8\linewidth]{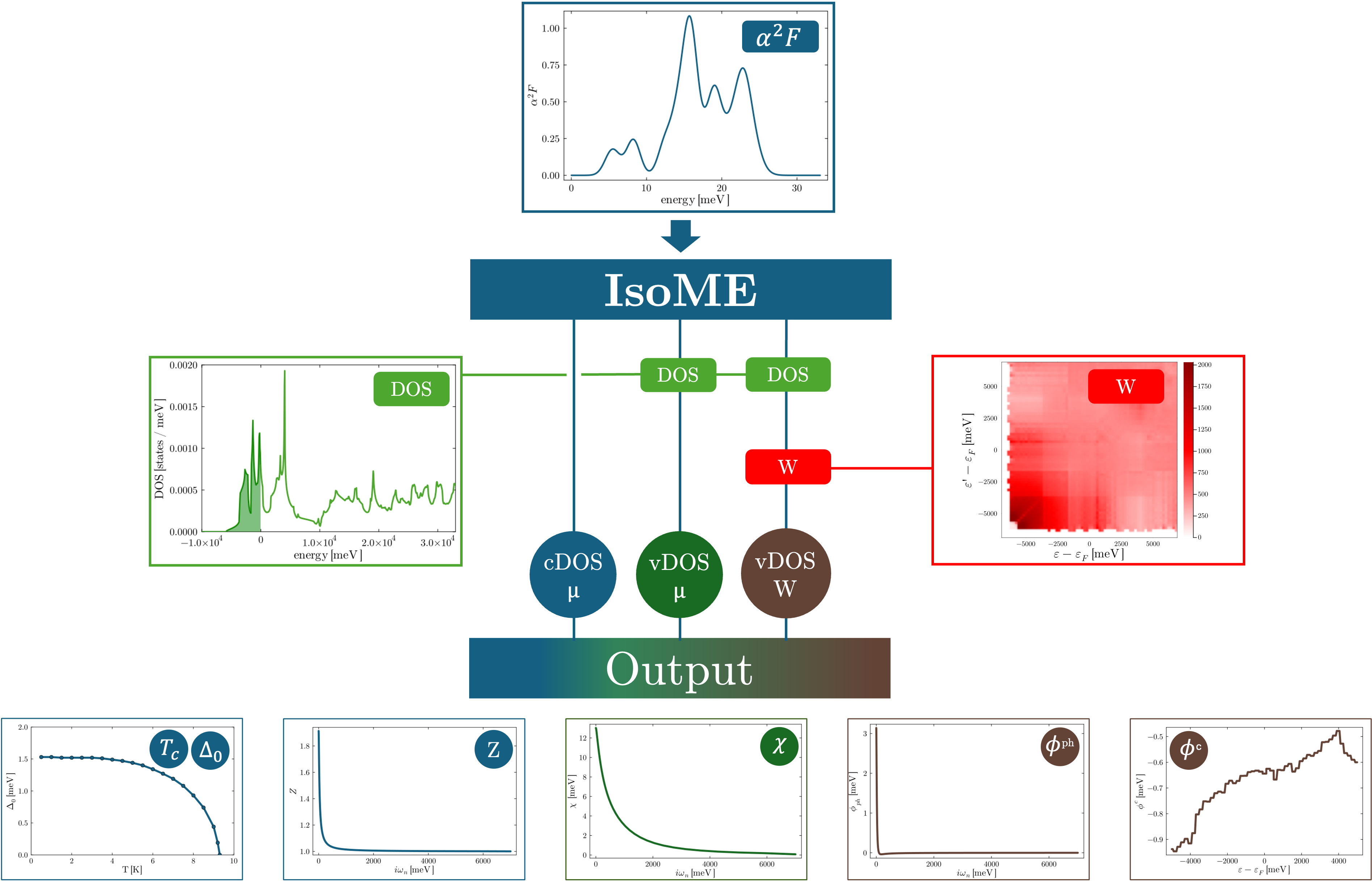}
    \caption{Flowchart of the \IsoME{} code showing how input and output files are associated with each level of theory. The $\afo$ file is the only input needed for the constant DOS (cDOS) approximation and is also required for all other computations. The electronic DOS and screened Coulomb interaction are specifically needed for variable DOS (vDOS) and W calculations, respectively. As output, \Tc{} and the components of the self-energy are returned.}
    \label{fig:IsoME_flowchart}
\end{figure*}

With this work, we aim to further facilitate the prediction and description of superconductors by introducing the very efficient and highly accurate, open-source Julia~\cite{Julia} package \IsoME{}~\cite{note_isoME}. \IsoME{} enables the calculation of superconducting properties at different levels of approximation within the framework of isotropic ME theory, as schematically depicted in Fig.~\ref{fig:IsoME_flowchart}. In its default setup, \IsoME{} solves the isotropic ME equations with just $\alpha^2F$ - the Eliashberg spectral function - 
as input, which can be obtained with modern density functional theory/density functional perturbation theory (DFT/DFPT) codes. For more advanced calculations, additional inputs such as the electronic density of states (DOS) and screened Coulomb interaction $W(\varepsilon, \varepsilon')$ can be incorporated, allowing for a fully \textit{ab-initio} treatment of superconductivity~\cite{lucrezi_full-bandwidth_2024, Sano-2016, sanna-2018, Wang2020, pellegrini_eliashberg_2022, Pellegrini2024}. Furthermore, \IsoME{} supports a \Tc{} search mode, eliminating the need for explicit temperature specification and enhancing efficiency in high-throughput calculations. Combined with the different levels of theory, this provides valuable information to guide experimental efforts.

The paper is structured as follows. In Sec.~\ref{sec:results}, we describe the functionalities of \IsoME{}, evaluate its performance, and present insights from our benchmarking study. Sec.~\ref{sec:method} reviews the ME theory and details the isotropic equations corresponding to the highest level of theory implemented in \IsoME{}. Finally, Sec.~\ref{sec:compdetails} provides the computational details of our calculations.

\section{Results and Discussion} \label{sec:results}

\subsection*{\IsoME{}}
The \IsoME{} code package~\cite{note_isoME} is highly flexible, enabling detailed analysis across the following multiple levels of approximation for studying the properties of the superconducting phase, schematically depticted in Fig.~\ref{fig:IsoME_flowchart}.

The set of equations \eqref{eq:me_Z}-\eqref{eq:me_phiph}, \eqref{eq:me_phic2}, and \eqref{eq:me_numel2}, as detailed in Sec.~\ref{sec:method}, constitutes currently the most complete and rigorous implementation to solve the isotropic Migdal-Eliashberg theory completely from first-principles, taking into account the full energy dependence of the electronic DOS and the screened Coulomb interaction, giving access to not only \Tc, but also to the superconducting gap function $\Delta$, the mass renormalization $Z$, the energy shift $\chi$, and thus the full Nambu-Gor'kov Green's function $\hat{G}$~\cite{Nambu}. Throughout this text, we will refer to this level of theory as vDOS+W (for \textit{variable DOS with static Coulomb $W(\varepsilon, \varepsilon')$}). A future version of \IsoME{} currently in development will allow to perform the transformation from Matsubara to real frequency space using Nevanlinna analytic continuation~\cite{khodachenko}.

The \IsoME{} package further allows to efficiently solve two simplified forms of Eliashberg theory, employing two common approximations.

\textit{(i) $\mu$ approximation}: 
The computation of $W$ is a challenging and resource-intensive task in itself. For cases where this is not possible or desired, we have implemented the option to use a single scalar $\mu^*$ - the Morel-Anderson pseudopotential~\cite{morel_anderson_1962} - instead of the full energy-dependent $W(\varepsilon, \varepsilon')$. 
It is defined as
\begin{equation}\label{eq:must}
    \mu^* = \frac{\mu}{1+\mu \text{ ln}\left(\frac{\varepsilon_\text{el}}{\hbar \omega_\text{ph}}\right)},
\end{equation}
where $\mu$ is the Fermi-surface averaged $W_{n\bbk,n'\bbkk}$ with band index $n$ and wavevector $\bbk$, $\omega_\text{ph}$ is a characteristic cutoff frequency for the phonon-induced interaction, and $\varepsilon_\text{el}$ is a characteristic electronic energy scale. When used within the McMillan-Allen-Dynes formula and its modifications, a \mustAD\ between 0.1 and 0.16 has been found to be appropriate for most conventional superconductors when compared to experiments~\cite{AllenDynes1975}.
Within ME theory, however, the corresponding phonon energy scale is given by the Matsubara frequency cutoff $\wc$, thus $\mu^*$ within the Eliashberg formalism (\mustE) needs to scale accordingly with $\wc$~\cite{Pellegrini2024}.

Within \IsoME{}, users can specify either $\mu$, \mustAD, or \mustE. When $\mu$ is provided, Eq. \eqref{eq:must} is used to calculate both \mustAD{} and \mustE. In case \mustAD{} is supplied, the corresponding \mustE{} is determined using~\cite{Pellegrini2024}
\begin{equation} \label{eq:mu*_e}
    \frac{1}{\mu^*_\text{E}}=\frac{1}{\mu^*_\text{AD}}+ \text{ln} \left( \frac{\omega_\text{ph}}{\wc} \right).
\end{equation}
Employing this approximation results in a simplified set of equations, as provided in [Eqs.~(1)-(6)] of the Supplemental Material (SM)~\cite{SM}. Throughout the paper, this approximation will be abbreviated as vDOS+$\mu$.

\textit{(ii) cDOS approximation}: 
\IsoME{} also supports the constant DOS (cDOS) approximation in the infinite bandwidth limit, in which case the Eliashberg equations simplify significantly as $\chi = 0$ and the chemical potential $\cp$ is constant. The corresponding set of equations is provided in Eqs. (7) and (8) of the SM~\cite{SM}. We will refer to this approximation as cDOS+$\mu$, implicitly assuming the infinite bandwidth approach when using the term cDOS.
At this lowest level of theory, illustrated by the blue color in the flowchart, only the Eliashberg spectral function $\afo$ is required as input, making \IsoME{} suitable for integration into high-throughput workflows. To further support such applications, we have implemented a dedicated \Tc{} \textit{search mode} for all levels of theory. This mode automatically determines \Tc{} to the nearest Kelvin without the need for explicit temperature specification.

We also want to mention at this point that considerable effort has been invested in selecting default parameters that, in most cases, ensure both computational efficiency and robust convergence. Further details can be found in Sec.~\ref{sec:method}.

\IsoME{} can be installed with the Julia package manager. Computations are started by handing over the input arguments to the \mbox{\textit{EliashbergSolver()}} function. As mentioned above, for the cDOS+$\mu$ approximation, only the $\afo$ file is required.

\scriptsize
\begin{lstlisting}[language=Julia]
 using IsoME

 inp =  arguments(
            a2f_file = "path-to-a2f-file",
            outdir   = "path-to-output-directory",
            )

 EliashbergSolver(inp)
\end{lstlisting}
\normalsize

All other input flags have predefined default values which are contained within the custom data type \textit{arguments()}. For further information, we refer to the documentation of the package. We want to stress at this point that \IsoME{} can automatically detect the different file formatting for \textsc{Quantum Espresso (QE)} ~\cite{QE-2009, QE-2017, QE-2020}, \textsc{EPW}~\cite{EPW-2013,EPW-2023} and \textsc{BerkeleyGW}~\cite{PhysRevB.34.5390, PhysRevB.62.4927, DESLIPPE20121269}. Automatic compatibility with other DFT/DFPT/$GW$ packages is currently under development. 

For more advanced calculations, the electronic DOS can also be provided, resulting in the vDOS+$\mu$ variant, depicted by the green color in Fig.~\ref{fig:IsoME_flowchart}.
The most rigorous approach - vDOS+W - incorporates both the variable DOS (vDOS) and the full static Coulomb interaction $W(\varepsilon,\varepsilon')$, as presented on the right-hand side of the flowchart. 
To initiate any of these variants, it is necessary to include the additional input files and to set the respective flags. 

\scriptsize
\begin{lstlisting}[language=Julia]
 using IsoME

 inp = arguments(
            a2f_file     = "path-to-a2f-file",
            dos_file     = "path-to-dos-file",
            Weep_file    = "path-to-W-file",
            outdir       = "path-to-output-directory",
            cDOS_flag    = 0,  # 0 = vDOS, 1 = cDOS
            include_weep = 1,  # 0 = mu, 1 = W
            )

 EliashbergSolver(inp)
\end{lstlisting}
\normalsize
\FloatBarrier

\begin{figure}[t]
    \centering
    \includegraphics[width=0.8\linewidth]{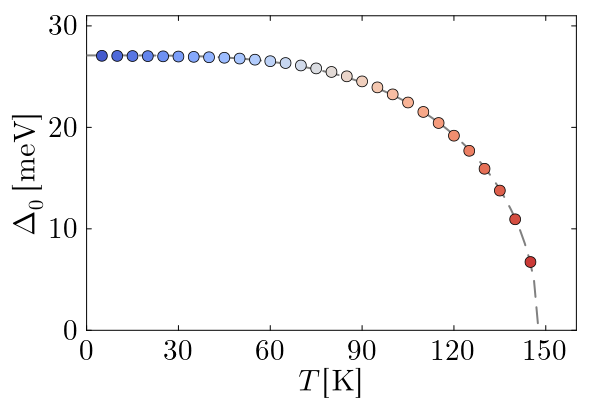}
    \caption{Temperature evolution of the superconducting gap $\Delta_0$ for LaBeH$_8$ as computed by \IsoME{}. The markers indicate the actual calculation points, and the dashed curve represents the fitted function described in the main text.}
    \label{fig:Gap}
\end{figure}

The outputs produced by \IsoME{} include the critical temperature \Tc, the superconducting gap function $\Delta$, and the renormalization function $Z$. When using the vDOS+$\mu$ or vDOS+W approaches, the energy shift $\chi$ and the phononic/electronic contributions to the anomalous self-energy $\phi^{ph/c}$ are also provided (see the bottom panel of Fig.~\ref{fig:IsoME_flowchart}). 
Moreover, \IsoME{} automatically generates a plot of the gap function at the zeroth Matsubara frequency, $\Delta_0$, which approximates the gap on the real frequency axis. Fig.~\ref{fig:Gap} presents the temperature dependence of $\Delta_0$ for LaBeH$_8$: the individual dots correspond to the calculated temperature points, while the dashed line serves as a guide to the eye and is the result of a fit using a functional form inspired by the BCS gap shape \cite{Gross1986},
$ \Delta_0(T) = \Delta_0(0) \, \tanh\!\left( \frac{\pi \kB\, T_\text{c}}{\Delta_0(0)} \sqrt{a\left(\frac{T_\text{c}}{T} - 1\right)} \right)$, with fitting parameters $a$, $\Delta_0(0)$, and \Tc{}.
Regardless of the selected approximation, the package also evaluates the critical temperature using both the McMillan-Allen-Dynes formula (Supplementary Eq.~(17)~\cite{McMillan2968, Dynes1972, AllenDynes1975}) and its machine learning-enhanced variant (Supplementary Eq.~(21)~\cite{Xie2022}).

\subsection*{Benchmarking}

Offering the above-mentioned range of approximations, \IsoME{} maintains high efficiency and performance.
Calculations can be performed on standard PCs within a reasonable timeframe, as shown in Tab.~\ref{tab:convergence_time}, which lists the runtimes for single-temperature calculations at 1, 10, and 100\,K for the hydride superconductor LaBH$_8$~\cite{Cataldo-labh8-2021} across the three theoretical levels available in \IsoME. 
Computation time grows as temperature decreases, due to the increasing number of Matsubara frequency points at lower temperatures. We implemented a sparse-sampling scheme ~\cite{shinaoka_compressing_2017, li_sparse_2020, WALLERBERGER2023101266}, to speed up the calculations at very low temperatures. 

\begin{figure*}[t]
    \centering
    \includegraphics[width=1\linewidth]{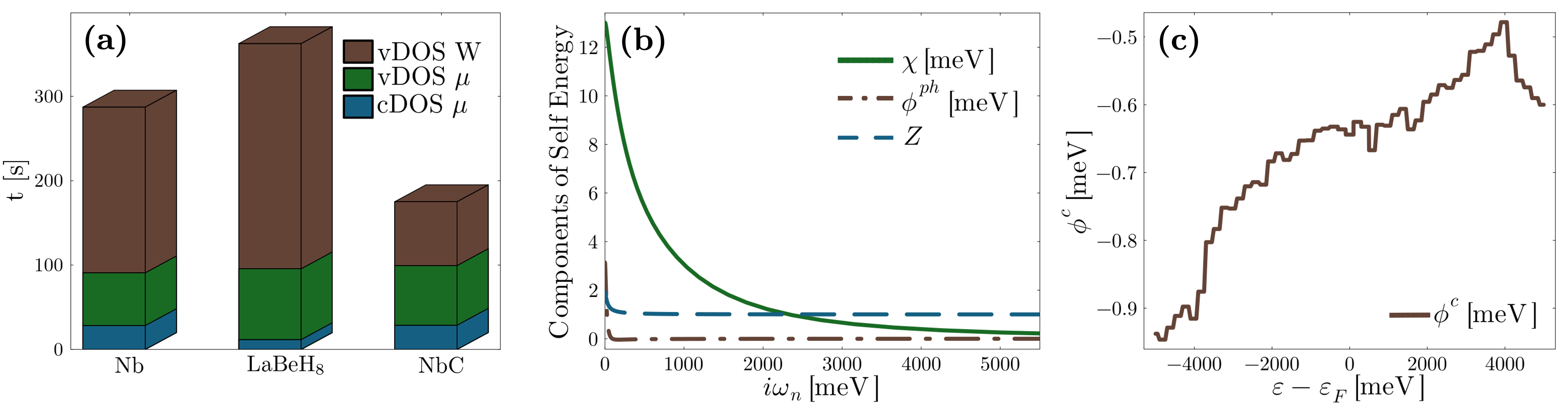}
  \caption{Duration of \Tc{} calculations and self-energy components of Nb. Panel (a) illustrates the runtime for the \Tc{} search mode for the different levels of theory for Nb, LaBeH$_8$, and NbC.
  In panel (b), the energy shift $\chi$ (solid green line), the renormalization function $Z$ (dashed blue line), and the phononic part of the order parameter $\phi^{ph}$ (dash-dotted brown line) are shown. The electronic part of the order parameter $\phi^c$ is displayed in panel (c) as solid brown line.}
  \label{fig:Nb_selfen}
\end{figure*}

Furthermore, \IsoME{} features an automatic \Tc{} search mode, which further enhances computational efficiency. Fig.~\ref{fig:Nb_selfen}a illustrates the runtime required to determine \Tc{} for Nb, LaBeH$_8$ and NbC. In the case of Nb, \IsoME{} calculated the \Tc{} within 28\,s using the cDOS+$\mu$ method, within $91$\,s with the vDOS+$\mu$ method, and within $287$\,s with the vDOS+W method. For LaBeH$_8$, the vDOS+$\mu$/W calculations took slightly longer than for Nb due to the broader temperature search range.
All these calculations were performed on a single core of an off-the-shelf workstation.

\begin{table}[t]
\centering
\caption{Computation runtimes of \IsoME{} for LaBH$_8$ at three temperatures for the three theoretical levels.}
\label{tab:convergence_time}
\begin{tabular}{c|r|r|r}
 & 1\,K &  10\,K &  100\,K \\
\midrule
cDOS+$\mu$ &  138\,s &   0.8\,s &    0.1\,s \\
vDOS+$\mu$ & 346\,s & 10.7\,s & 1.9\,s \\
vDOS+W &  370\,s &  28.0\,s &   15.0\,s \\
\bottomrule
\end{tabular}
\end{table}

\begin{table*}[t]
    \centering
    \caption{Comparison of resulting \Tc~for different levels of theory within \IsoME. \TcAD, \TcML, \TcEcdosmu, \TcEvdosmu, \TcEvdosW, and \Tcexp~are the critical superconducting temperatures within McMillan-Allen-Dynes, the machine learned improvement of it, Migdal-Eliashberg constant DOS and variable DOS using $\mu$ as input, Migdal-Eliashberg variable DOS including static Coulomb interactions $W(\epsilon, \epsilon')$, and the experimental value, respectively. The given $\mu$ value was calculated within $GW$. Based on $\mu$, two seprate values for $\mu^*$ are determined based on Eq. \eqref{eq:must}. \mustAD{} to be used for \TcAD{} and \TcML{} with $\omega_\text{ph}$ being the highest phonon frequency of the system and \mustE{} to be used for \TcEcdosmu{} and \TcEvdosmu{} with $\omega_\text{ph}$ corresponding to the Matsubara cutoff $\wc$. For reference, \mustAD{} and \mustE{} are listed explicitly in Tab. II of the SM~\cite{SM}. The asterisk indicates alloyed materials, where the crystal structure and 1:1 stoichiometry used to model the material may differ from the experimental one.}

    \label{tab:tc_5000}
    \setlength{\tabcolsep}{6pt}
	    \begin{tabular}{c|ccccccc}
    \toprule
         Compound & $\mu$ (from $W$) & \TcAD &  \TcML &  \TcEcdosmu &  \TcEvdosmu &\ \TcEvdosW & \Tcexp \\
                  & & [K] &[K] &[K] &[K] &[K] &[K] \\
         \midrule
         Nb & 0.38 & 9  & 9  & 9  & 8  & 7 &  9.1-9.5~\cite{matthias1963superconductivity} \\  
         Al & 0.26 & 1 & 1 & 1 & 1 &     1 & 1.2~\cite{matthias1963superconductivity}          \\
         Tc & 0.21 & 16  & 17  & 16  & 16   & 16 & 8.2-9.3-11.1~\cite{matthias1963superconductivity}     \\
         NbC & 0.17 & 17  & 17  & 16  & 15    &  16 &  12.8*~\cite{giorgi1962}         \\
         TiN & 0.13 & 18  & 17  & 17  & 16   & 17 &   5.6*~\cite{matthias1963superconductivity}         \\
         H$_3$S (200\,GPa) & 0.29 &225& 255& 237& 216 & 211 & 172-184 ~\cite{drozdov_conventional_2015} \\
         YH$_6$ (200\,GPa) & 0.20 &   208 & 237 & 231 &  226  & 227   &   208-214~\cite{kong2021superconductivity}   \\
         LaBeH$_8$ (100\,GPa) &0.16 &130&  139 & 141  &   140 &  140  &  104~\cite{SongLaBeH8-exp2023}    \\
         \labh (50\,GPa) & 0.13 & 147   &  161   &  152   & 154     &  158  &   -     \\
         \bottomrule
    \end{tabular}
\end{table*}

To quantitatively validate our computational framework, we first performed benchmark calculations of \Tc's (to the nearest Kelvin) for an extensive set of materials using the default value of \mustAD{} = 0.12. This approach allowed us to sidestep the considerable computational cost associated with a full $W$ calculation. The results, presented in Tab.~I of the SM~\cite{SM}, show excellent agreement across the various levels of approximation and with experimental data.

Encouraged by these promising benchmarks, we subsequently computed both $W$ and $\mu$ using \textsc{BerkeleyGW}~\cite{PhysRevB.34.5390,PhysRevB.62.4927,DESLIPPE20121269} for several particularly interesting materials, thereby obtaining fully \textit{ab initio} results from ME theory. The set of materials comprises elementary, binary, and ternary superconductors, including high-pressure hydrides. The results are summarized in Tab.~\ref{tab:tc_5000}. 

\textit{Elementary superconductors}: 
Elementary superconductors typically consist of only a few atoms per unit cell, making them amenable to both detailed experimental characterization and accurate theoretical modeling. This is confirmed by our benchmark study on Nb, Al, and Tc, where all approximations yield very similar \Tc{} values. For instance, 
when comparing calculations with constant and variable DOS for Nb - which exhibits a pronounced peak at the Fermi level - we observe a deviation of only 1\,K. Furthermore, incorporating an energy-dependent Coulomb interaction $W(\varepsilon,\varepsilon')$ does not notably alter the computed \Tc{} for these materials. Overall, our results agree closely with experimental measurements, with one notable exception: Technetium. Here, reproducing the experimental \Tc{} of 11\,K would require a larger effective Coulomb parameter than the calculated static value of $\mu=0.21$. This discrepancy is consistent with the findings of Ref.~\cite{kawamura2020benchmark}, which suggest that spin fluctuations - absent in our Coulomb-only treatment - may further suppress \Tc{} in Technetium.

While the semi-empirical equations provide accurate estimations for \Tc{}'s of elementary superconductors, they do not provide any insight into the nature of the superconducting state. ME theory, on the other hand, offers a detailed description of the self-energy components, which we show as an example for Nb in panels (b) and (c) of Fig. \ref{fig:Nb_selfen}. The following characteristics are observed across all materials considered:
At moderately high Matsubara frequencies, the mass renormalization function $Z$ quickly approaches unity, while the phonon contribution to the order parameter, $\phi^{ph}$,  vanishes. In contrast, the energy shift $\chi$ decays to zero only at significantly higher frequencies. Moreover, the electronic component of the order parameter, $\phi^c$, exhibits distinct features that reflect the influence of both the DOS and the energy-dependent Coulomb interaction $W(\varepsilon, \varepsilon')$.
 
\textit{Disordered binary alloy superconductors}: Alloy superconductors, including NbN, NbTi, TiN, and NbC, are of particular interest due to their prevalence in numerous technological applications. However, unlike elementary superconductors, modeling disordered binary superconductors poses significant challenges, often attributable to vacancies that result in an off-stoichiometry phase rather than a 1:1 ratio. 
To benchmark \IsoME{}, we approximated the structure of the alloys by a 1:1 stoichiometry, which leads to an overestimation of \Tc{} across all levels of approximation. The effect is particularly pronounced for TiN, where the calculated \Tc{} is three times higher than the experimental one.  For NbC, the calculated \Tc{} aligns more closely with experiments, particularly when using a standard \mustAD\ value of 0.12 instead of the computed one (see Tab.~I in the SM~\cite{SM}).
For alloyed systems, the limitation arises not from the theoretical treatment of superconductivity itself but from the inadequate representation of the disordered crystal structure. 
To accurately capture defects, large supercells are necessary, rendering simulations prohibitively costly and time-consuming. Methods such as the supercell approach used for NbTi in Ref.~\cite{Cucciari-2024} or the ECQCA framework \cite{FERREIRA2024} provide potential solutions, yet these refinements are beyond the scope of this work.

\textit{Hydrides}: 
Building on Ashcroft's prediction that atomic hydrogen under high pressure could have a record-breaking high \Tc{}~\cite{Ashcroft}, hydrogen-based superconductors have attracted significant attention. Some compounds achieve superconductivity at or near room temperature but at unfeasibly high pressures for practical applications. Considerable efforts are being made to realize a hydrogen superconductor stable under near-ambient conditions. Modeling such materials accurately is particularly challenging, as hydrogen compounds can exhibit properties distinct from other conventional superconductors, including exceptionally strong electron-phonon coupling, unique electronic features at the Fermi level, and quantum anharmonic effects~\cite{Boeri_JPCM_2021_roadmap}.

Indeed, the hydrides included in our calculations exhibit the most pronounced discrepancy between different levels of theory. 
In contrast, for non-hydrides, both the McMillan-Allen-Dynes equation and its machine-learned improved variant agree with ME theory within 2\,K. However, hydrides differ from the materials used to fit the AD~equation, leading to deviations of 19, 10, and 11\,K for YH$_6$, LaBeH$_8$, and LaBH$_8$ respectively, when compared to the vDOS+W approximation, while the \TcML{} model aligns with ME calculations within a few percent. Interestingly, for \hs, the difference for \TcAD{} is 19\,K, whereas the machine-learned formula deviates by 44\,K from \TcEvdosW.
This underscores the importance of \textit{ab-initio} methods to predict superconductors, as novel materials may exhibit properties not (well) represented in certain datasets. Additionally, \textit{first-principles} methods give more insights into the underlying physics, such as access to the gap function $\Delta$ and other components of the self-energy (see Fig. \ref{fig:Gap} and \ref{fig:Nb_selfen}).

It is important to note that perfect agreement between calculated and experimental \Tc{} values for hydride superconductors is not expected. Our calculations are based on ideal, high-symmetry structures with perfect three-dimensional periodicity (i.e., single crystals). In contrast, experimental syntheses - especially under extreme conditions - often yield samples with impurities, vacancies, and other types of disorder that can alter superconducting properties. Moreover, quantum anharmonic effects, which are particularly significant in these materials, are not included in our current approach~\cite{PhysRevLett.114.157004, Errea_2020, Lucrezi2023, Lucrezi2024}. Despite these differences, the strong agreement between our calculated \Tc{} values and experimental observations underscores the robustness of our theoretical framework.

\begin{figure*}[t]
    \centering
    \includegraphics[width=1\linewidth]{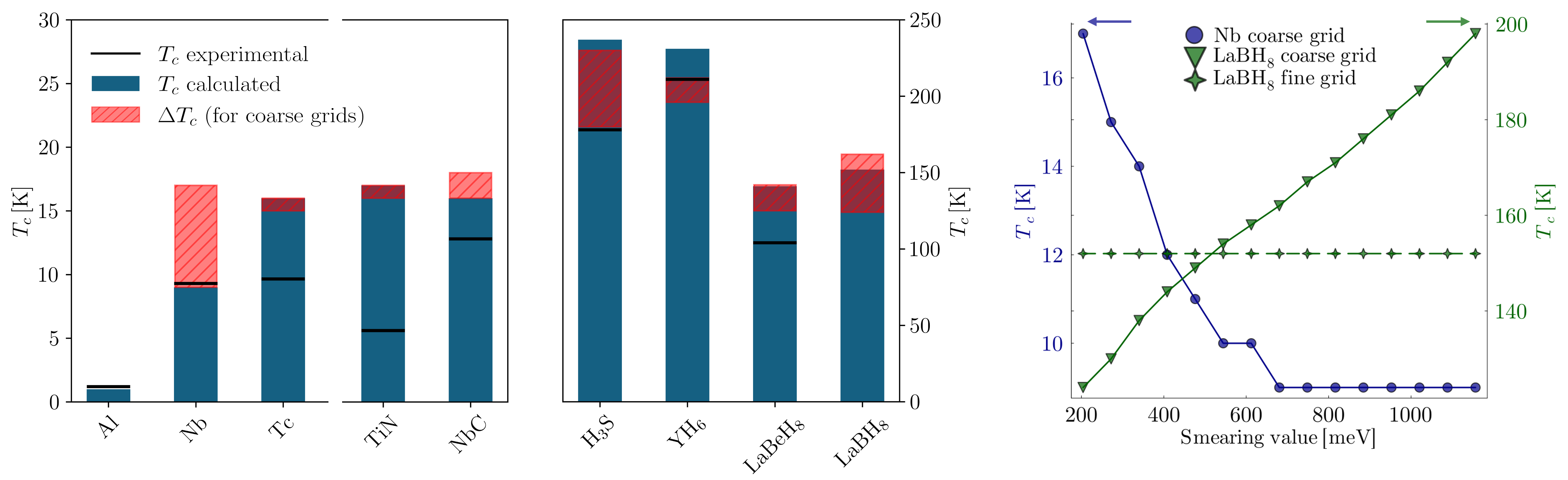}
    \caption{Comparison of \Tc{} values for different smearing values for materials in the benchmark study. Left: The calculated \Tc{}'s with cDOS+$\mu$ are shown as blue bars, the experimental values as black lines. For LaBH$_8$, no experimental data is available. Red and crosshatched bars indicate the \Tc{} values in the unconverged region. 
    Right: Convergence behavior of Nb (blue dots) and LaBH$_8$ (green triangles and stars) with respect to smearing. The lines serve as guides to the eye.}
    \label{fig:smearing}
\end{figure*}

The results obtained for constant or variable DOS using the $\mu$ approximation differ only by a few percent for YH$_6$, LaBeH$_8$, and LaBH$_8$. \hs, however, shows a pronounced peak at the Fermi level, where the assumption of a constant DOS around the Fermi level is no longer valid. As a result, a difference of 21\,K is observed, where the \Tc{} calculated using vDOS+$\mu$ is closer to the experimental value.  
The largest deviation in \Tc{} between vDOS+$\mu$ and vDOS+W occurs for H$_3$S and LaBH$_8$, with differences of 5 and 4\,K, respectively. In LaBH$_8$, the difference is due to the strong variation of $W(\varepsilon, \varepsilon')$ near the Fermi energy, whereas in H$_3$S, it results from the pronounced peak in the electronic DOS at the Fermi level, which directly enters Eq.~\ref{eq:me_phic2}.
In general, taking into account the full energy-dependent Coulomb repulsion $W(\varepsilon, \varepsilon')$, instead of reducing it to a scalar $\mu$ had little influence on most materials in our benchmark study, as can be seen by comparing the columns labelled \TcEvdosmu{} and \TcEvdosW{} in Table~II.
 However, a more pronounced impact is expected in Chevrel phases, multigap superconductors, or low dimensional systems \cite{pellegrini_eliashberg_2022}. 

\IsoME{} was designed as a robust and user-friendly framework for calculating superconducting properties. Nevertheless, for reliable \Tc{} predictions and proper interpretation of the results, several aspects must be considered when using \IsoME{}:

First, accurate results can only be achieved through carefully conducted convergence tests for the input files. In particular, the $\afo$ data needs to be of sufficient quality. 
In the following, we examine how different Brillouin zone grids influence \Tc.
The electron-phonon coupling for most materials in Tab.~\ref{tab:tc_5000} was computed with \textsc{Quantum Espresso}, where very dense \textit{q}-grids can get prohibitively expensive, requiring smearing techniques to facilitate convergence.
However, as Fig.~\ref{fig:smearing} illustrates, varying the smearing parameters in the $\afo$ calculations can lead to significant deviations in the predicted \Tc. In several instances (e.g., \hs, YH$_6$, LaBeH$_8$, and LaBH$_8$), convergence could not be attained with the \textsc{Quantum Espresso} $\afo$ data. To overcome this limitation, we interpolated the data to a significantly denser grid using \textsc{EPW}, effectively eliminating the dependence on smearing. The right panel of Fig.~\ref{fig:smearing} further illustrates this behavior: while Nb attains convergence even with coarser grids for larger smearing values, LaBH$_8$ fails to converge under these conditions. Although these results were obtained using the cDOS+$\mu$ approximation, we observed a consistent trend across different theoretical approaches.

Second, the choice of $\mu$ significantly influences the results. Traditionally, $\mu^*$ is treated as an adjustable parameter and is typically chosen within the range of 0.1 to 0.16 to fit experimental values. 
For fully \textit{ab-initio} calculations, $\mu$ must be computed and \mustE{} adapted according to Eq.~\eqref{eq:mu*_e}, as was done for the results presented in Tab.~\ref{tab:tc_5000}.

In light of these considerations, we compare the results presented in Tab.~\ref{tab:tc_5000} with those available in the literature and reexamine YH$_6$ and H$_3$S, which were also studied by some of the present authors in Ref.~\cite{lucrezi_full-bandwidth_2024}. First, a comment on the semi-empirical approximations: In the present work, the McMillan formula modified by Allen-Dynes was applied, whereas in Ref.~\cite{lucrezi_full-bandwidth_2024}, the original McMillan equation as given in Eq.~(9) in the SM~\cite{SM} was used together with a \mustAD~of~0.1. If we employ the McMillan equation and the same \mustAD{} for YH$_6$ and H$_3$S, we obtain the previously reported \Tc's of 154 and 173~K, respectively. Additionally, also \TcML~is reproduced when setting ${\text{\mustAD} = 0.1}$. 
Second, comparing \Tc's based on ME theory is less straightforward. While our results are based on the isotropic approximation, the findings in Ref.~\cite{lucrezi_full-bandwidth_2024} are obtained from anisotropic calculations. To ensure a meaningful comparison, we performed \IsoME{} calculations using the same parameters as reported in Ref.~\cite{lucrezi_full-bandwidth_2024}, including a Matsubara frequency cutoff $\wc$ of 6,000\,meV, an electronic energy cutoff of 1,000\,meV, and an unmodified \mustE. Under these conditions, our results across all approximations deviate by no more than 7\% for YH$_6$ and 5\% for H$_3$S.

\section*{Conclusion}
We have developed the open-source Julia package \IsoME{}, which provides a straightforward and efficient method for solving the isotropic Migdal-Eliashberg equations and evaluating \Tc{} across different levels of approximation. These range from the semi-empirical McMillan-Allen-Dynes formula and the ML variant to the isotropic Migdal-Eliashberg approach, which can include either a constant or variable DOS and static Coulomb interactions. To demonstrate the accuracy and versatility of the code, we conducted a benchmark study on a diverse set of materials, including elementary, binary alloy, and ternary hydride superconductors, and compared the results to existing literature. Although the constant DOS approximation works well for many conventional superconductors, \IsoME{}’s flexible framework also supports the analysis of systems with complex electronic structures, which require both a refined treatment of the DOS near the Fermi level and the explicit inclusion of Coulomb interactions.

\IsoME{} represents a significant advancement in the computational treatment of superconductivity, by offering multiple levels of approximation and a dedicated \Tc{} search mode. This makes it a valuable tool for both theoretical investigations and high-throughput material discovery, paving the way for more accurate and efficient predictions of superconducting properties.

\section{Methods} \label{sec:method}
\subsection*{Isotropic Midgal-Eliashberg theory}

Conventional superconductivity arises from the formation of Cooper pairs, where an attractive interaction between electrons mediated by the electron-phonon coupling enables them to overcome the repulsive Coulomb interaction. 
This is described by ME theory, most conveniently formulated within the Nambu-Gor'kov formalism~\cite{Nambu}, leading to a generalized $2 \times 2$ Green's function propagator in Matsubara frequency space given by
\begin{equation}
\label{eq:MatrixPropagator}
    \hat{G}_{\nk}(i\wj) = 
    \begin{pmatrix}
    G_{\nk}(i\wj) & F_{\nk}(i\wj)  \\
    F_{\nk}^*(i\wj) & -G_{n-\mathbf{k}}(-i\wj) \\
    \end{pmatrix} ~.
\end{equation}
The diagonal elements describe single particle excitations, the off-diagonal elements are the anomalous propagators describing the Cooper-pair amplitude, with band index $n$, wavevector $\bbk$, and $i\wj$ as fermionic Matsubara frequencies.  
The Dyson equation relates the interacting ($G_{\nk}$) and non-interacting Green's functions ($G_{\nk}^0$) via a self-energy $\hat{\Sigma}_{\nk}(i\wj)$~\cite{Ponc_2016, mattuck_guide_1992} 
\begin{equation}
    \hat{G}_{\nk}^{-1}(i\wj) = \hat{G}_{\nk}^{0^{-1}}(i\wj) - \hat{\Sigma}_{\nk}(i\wj) ~,
\end{equation}
which can be further decomposed in Nambu space using the Pauli matrices $\tau_i$~\cite{ScalapinoSchrieffer}:
\begin{equation}
\label{eq:self_e}
\begin{split}
    \hat{\Sigma}_{\nk}(i\wj) = &i \wj \left[ 1 - Z_{\nk}(i\wj) \right] \hat{\tau}_0 \\
    &+ \chi_{\nk}(i\wj)\hat{\tau}_3 + \phi_{\nk}(i\wj)\hat{\tau}_1~.
\end{split}
\end{equation}
The as of yet unknown components of the self-energy can be identified as the mass renormalization function $Z$, the energy shift $\chi$, and the anomalous self-energy $\phi$~\cite{lucrezi_full-bandwidth_2024}.

On the other hand, based on the Fröhlich Hamiltonian for a coupled electron-phonon system with Coulomb interactions and employing Migdal's approximation~\cite{migdal1958,eliashberg1960,ALLEN19831}, we can find an alternative expression for the electron self-energy $\hat{\Sigma}_{\nk}(i\omega_{n})$ given by 
\begin{equation}
\label{Sigma_1}
\begin{split} 
	\hat{\Sigma}_{\nk}(i\wj) 
	= -\frac{1}{\beta} &\sum_{\bbkk n'j'} \hat{\tau}_3 \hat{G}_{\npkp}(i\wjp) 
	\hat{\tau}_3 \bigg[ \Wnkwj \\ 
    + &\sum_{\lambda} |g_{nn'\lambda}(\bbk,\bbkk)|^2 D_{\bbk-\bbkk\lambda}(i\wj-i\wjp) \bigg]~.
\end{split}   
\end{equation}
Here, the first term describes the dynamically screened vertex-corrected Coulomb electron-electron interaction \mbox{$\Wnkwj$}, while the second term describes the interaction of electrons and phonons (with mode index $\lambda$) via the electron-phonon coupling matrix element $g_{nn'\lambda}(\bbk,\bbkk)$ and the phonon propagator $D_{\bbk-\bbkk\lambda}(i\omega_{n}-i\wjp)$ ~\cite{marsiglio2001, ALLEN19831}. The inverse temperature is denoted by $\beta=(\kB T)^{-1}$.

Assuming a static Coulomb interaction~\cite{Pellegrini2024} and using the spectral representation of $D_{\bbk-\bbkk\lambda}(i\omega_{n}-i\wjp)$, the electron self-energy from Eq.~\eqref{Sigma_1} becomes
\begin{equation}
\label{eq:Sigma_a2f}
\begin{split} 
	\hat{\Sigma}_{\nk}(i\wj) 
	= &-\frac{1}{\beta} \sum_{\bbkk n'j'} \hat{\tau}_3 \hat{G}_{\npkp}(i\wjp) 
	\hat{\tau}_3 \, \times \\ 
    &\Bigg[ \Wnk -\int_0^{\infty} \!\mathrm{d}\omega \frac{\afkko}{\NF} \frac{2\omega}{(\wj-\wjp)^2+\omega^2} \Bigg]~,
\end{split}   
\end{equation}
where we have introduced the Eliashberg spectral function~\cite{ALLEN19831, marsiglio2001}
\begin{equation}
	\afkko = \NF \sum_{\lambda} |g_{nn'\lambda}(\bbk,\bbkk)|^2 B_{\bbk-\bbkk\lambda}(\omega)~,
	\label{a2F_kko_1}
\end{equation}
and the density of states per spin at the Fermi level $\ef$, $\NF$.
Note: In most cases, the fully interacting phonon Green's function is well approximated by choosing \mbox{$ B_{\bbk-\bbkk\lambda}(\omega)  \approx \delta(\omega-\omega_{\bbk-\bbkk,\lambda})$}~\cite{ALLEN19831}.

Lastly, defining the electron-phonon coupling parameter as~\cite{AllenDynes1975, pickett_generalization_1982}
\begin{equation}
	\lambda_{\nk,\npkp}(\wj) 
	= \int_{0}^{\infty} \!\mathrm{d}\omega\ \afkko \frac{2\omega}{\wj^2+\omega^2}
	\label{lambda_3}
\end{equation}
allows us to rewrite Eq.~\eqref{eq:Sigma_a2f} in a more compact form 
\begin{equation} 
\label{eq:Sigma_lambda}
\begin{split}
	\hat{\Sigma}_{\nk}(i\wj) 
	= &-\frac{1}{\beta} \sum_{\bbkk n'j'} 
	\hat{\tau}_3 \hat{G}_{\npkp}(i\wjp) \hat{\tau}_3 \\
	&\times \left[W_{\nk,\npkp} -\frac{\lambda_{\nk,\npkp}(\wj-\wjp)}{\NF} 
	 \right]~.
 \end{split}
\end{equation}

By comparing the two different expressions for the self-energy in Eq.~\eqref{eq:self_e} and Eq.~\eqref{eq:Sigma_lambda}, we arrive at a set of coupled equations - the \textit{anisotropic full-bandwidth Migdal-Eliashberg equations} - that determine the previously introduced functions $Z,~\chi$ and $\phi$,  as discussed in [Eqs.~(13)-(16)] of Ref.~\cite{lucrezi_full-bandwidth_2024} and allow a very detailed, wave-vector resolved investigation of the properties of the superconducting phase.

Solving these equations is extremely demanding computationally, as the electron-phonon coupling can vary significantly near the Fermi surface, requiring very fine $\mathbf{k}$-point sampling. However, for most materials - except for a few notable layered systems such as MgB$_2$~\cite{Kortus-2001-MgB2, Choi-2002-MgB2, Choi-2003-MgB2}, CaC$_6$~\cite{Gonnelli-2008-CaC6, Sanna-2007-CaC6}, and  NbS$_2$~\cite{Heil2017} - this complexity can be mitigated by employing the isotropic approximation, where all quantities are averaged over the Brillouin zone~\cite{Davydov2020,Pellegrini2024}.

In particular, for the Eliashberg spectral function and the static Coulomb interaction we get
\begin{equation}
    \label{a2f_fermiSurface}
    \afo = \frac{1}{N(\ef)^2}\sum_{\nk,\npkp}\afkko\delta(\varepsilon_{\nk}-\ef)\delta(\varepsilon_{\npkp}-\ef)~
\end{equation}
\begin{equation}
    \label{eq:aver_W}
    W(\varepsilon, \varepsilon') = \frac{1}{N(\varepsilon)N(\varepsilon')}\sum_{\nk,\npkp}\Wnk\delta(\varepsilon_{\nk}-\varepsilon)\delta(\varepsilon_{\npkp}-\varepsilon), 
\end{equation}
respectively.

The final set of coupled equations for the self-energy components in the isotropic approximation, as implemented in \IsoME{}, is given by
\begin{widetext}
\begin{subequations}
\begin{align}
\label{eq:me_Z}
Z(i\wj) &=  1+ \frac{\kB T}{\NF\wj} \int\!\mathrm{d}\varepsilon' N(\varepsilon') 
\sum_{j'} \frac{\wjp Z(i\wjp)}{\Theta(\varepsilon', i\wjp)} \lambda(\wj-\wjp) \\
 \label{eq:me_chi}
    \chi(i\wj) &= - \frac{\kB T}{\NF} \int\!\mathrm{d}\varepsilon' N(\varepsilon') 
    \sum_{j'} \frac{\varepsilon'-\cp + \chi(i\wjp)}{\Theta(\varepsilon', i\wjp)} \lambda(\wj-\wjp)   \\
    \label{eq:me_phiph}
    \phi^{ph}(i\wj) &= \frac{\kB T}{\NF} \int\!\mathrm{d}\varepsilon' N(\varepsilon')   \sum_{j'} \frac{\phi(\varepsilon',i\wjp)}{\Theta(\varepsilon', i\wjp)} \lambda(\wj-\wjp) \\
    \label{eq:me_phic}
    \phi^c(\varepsilon) &= -\kB T \int\!\mathrm{d}\varepsilon' N(\varepsilon') W(\varepsilon, \varepsilon') \sum_{j'} \frac{\phi(\varepsilon',i\wjp)}{\Theta(\varepsilon', i\wjp)}  \\
    \label{eq:me_numel}
    N_e &= \int\!\mathrm{d}\varepsilon' N(\varepsilon') \left[ 1 - 2 \kB T \sum_j \frac{\varepsilon'-\cp + \chi(i\wj)}{\Theta(\varepsilon', i\wj)}\right],  
\end{align}
\end{subequations}
\end{widetext}
with $\Theta(\varepsilon, i\wj) = [ \wj Z(i\wj) ]^2 + [ \varepsilon - \cp + \chi(i\wj)]^2 + [ \phi(\varepsilon,i\wj)]^2$ and $\phi(\varepsilon,i\wj) = \phi^{ph}(\wj)+\phi^c(\varepsilon)$. A detailed derivation is not within the scope of this manuscript and can be found elsewhere~\cite{EPW-2023, pellegrini_eliashberg_2022}. The last  Eq.~\eqref{eq:me_numel} fixes the electron number and is used to determine the chemical potential $\cp$ self-consistently~\cite{lucrezi_full-bandwidth_2024, pavarini_emergent_2013}. From the anomalous self-energy $\phi$ and the renormalization function $Z$, the superconducting gap function can be obtained via
\begin{equation}
\label{eq:me_delta}
   \Delta (\varepsilon, i\omega_j) = \frac{\phi(\varepsilon,i\wj)}{Z(i\wj)}~.
\end{equation}
The critical superconducting temperature \Tc{} is defined as the temperature for which $\Delta$ vanishes. \\

Equation~\eqref{eq:me_delta} concludes the theoretical framework. In the following, we discuss numerical challenges and provide implementation details. In equations involving a sum over Matsubara frequencies, the summation formally extends over infinitely many frequencies. To perform such an infinite sum numerically, a cutoff frequency, $\wc$, is introduced as a convergence parameter. $\lambda$ decays as $1/\omega_j^2$ and thus $Z$, $\chi$, and $\phi^{ph}$ decay as $1/\omega_j^4$, ensuring convergence for relatively small $\wc$ (of the order of 10-20 times the Debye frequency of the system). This is not the case for $\phi^c$ and $N_e$, where considerably higher cutoffs are needed. A similar argument holds for the integrals over real energies $\varepsilon$, where electron-phonon interactions are restricted to energies close to the Fermi energy while Coulomb interactions decay considerably slower.
One can improve convergence and accuracy by approximating $Z(i\wj)=1,~\phi(\varepsilon,i\wj) = 0$ and $\chi(i\wj)=0$ for $\omega_j > \wc$ and $|\varepsilon| > \varepsilon_c$, allowing to rewrite Eqs.~\eqref{eq:me_phic} and \eqref{eq:me_numel} as~\cite{note_analytic_tail, pellegrini_eliashberg_2022} 

\begin{widetext}
\begin{subequations}
\begin{equation}
\label{eq:me_phic2}
\phi^c(\varepsilon) = \int\!\mathrm{d}\varepsilon' N(\varepsilon')W(\varepsilon, \varepsilon') \left\{ \frac{\phi^c(\varepsilon')}{2}  \frac{\tanh\left[\frac{\beta}{2}\sqrt{(\varepsilon'-\cp)^2 + \phi^{c^2}(\varepsilon')} \right] }{\sqrt{(\varepsilon'-\cp)^2 + \phi^{c^2}(\varepsilon')}} + 2\kB T \sum_{j'=0}^{\wj\leq\wc} \left[ \frac{\phi(\varepsilon',i\wjp)}{\Theta(\varepsilon', i\wjp)} - \frac{\phi^c(\varepsilon')}{\wjp^2+(\varepsilon'-\cp)^2+\phi^{c^2}(\varepsilon')} \right] \right\} 
\end{equation}
\begin{equation}
\label{eq:me_numel2}
    N_e \approx \int\!\mathrm{d}\varepsilon N(\varepsilon) \left\{ 2\nf(\varepsilon-\cp) - 4\kB T \sum_{j=0}^{\wj\leq\wc}\left[
    \frac{\varepsilon-\cp + \chi(i\wj)}{\Theta(\varepsilon,i\wj)} 
    -\frac{\varepsilon-\cp}{\wj^2 + \left(\varepsilon-\cp\right)^2} \right]\right\}~.
\end{equation}
\end{subequations}
\end{widetext}

To maintain charge neutrality in vDOS+$\mu$ and vDOS+W calculations, the chemical potential should be continuously updated according to Eq.~\eqref{eq:me_numel2}.

Within \IsoME{}, we found that choosing $\wc \sim 5–10$\,eV is sufficient to ensure both computational feasibility and high accuracy. Additionally, we implemented a sparse-sampling scheme to enhance the efficiency of the summation over Matsubara frequencies~\cite{shinaoka_compressing_2017, li_sparse_2020, WALLERBERGER2023101266}. Given that the overhead associated with setting up the irreducible sparse-sampling basis outweighs the time savings from the summation for temperatures above 2\,K, sparse-sampling is activated only for temperatures below this threshold.

\section{Computational Details}\label{sec:compdetails}

\subsection*{DFT and DFPT calculations}

Density functional (perturbation) theory calculations were performed using \textsc{Quantum Espresso}~\cite{QE-2009, QE-2017, QE-2020}, scalar-relativistic optimized norm-conserving Vanderbilt pseudopotentials~\cite{vanbilt_pseudo_hamann_2013_PhysRevB.88.085117}, and the PBE-GGA exchange and correlation functional~\cite{GGA_perdew_1996PhysRevLett.77.3865}. For the following materials, a plane-wave cutoff energy for the wavefunctions of 80\,Ry was applied, and the respective $\mathbf{k}$-grid for each material is given in parenthesis: Nb ($16^3$), Tc (18$^3$), Sn (8$\times$8$\times$14), TiN (12$^3$), NbC (14$^3$),  \hs~(20$^3$), YH$_6$ (16$^3$), and LaBeH$_8$~(12$^3$).
A cutoff of 50\,Ry and a $\mathbf{k}$-grid of 20$^3$ were used for Al. LaBH$_8$ was calculated with a cutoff of 90~Ry and a 16$^3$ $\mathbf{k}$-grid. All materials were calculated with a Methfessel-Paxton \cite{Methfesselpaxton} smearing of 0.02\,Ry, except YH$_6$, where a smearing of 0.04\,Ry was applied. 
For the relaxation of lattice parameters and atomic positions, the convergence threshold was set to 10$^{-7}$\,Ry and 10$^{-6}$\,Ry/a$_0$ for the total energy and the forces, respectively. DFPT calculations were conducted on a coarse 6$^3$ $\mathbf{q}$-grid with a self-consistency threshold of 10$^{-14}$ or lower for all materials. The fine grid consisted of 48$^3$ $\mathbf{k}$- and $\mathbf{q}$-points for YH$_6$ and H$_3$S, and 30$^3$ $\mathbf{k}$- and $\mathbf{q}$-points for LaBeH$_8$ and LaBH$_8$. More details on the interpolation to the fine grids are provided in \cite{lucrezi_full-bandwidth_2024, khodachenko}.

\subsection*{\IsoME{}}

For the calculations with \IsoME{}, a Matsubara frequency cutoff \textit{omega\_c} of 7,000\,meV was employed. In variable DOS calculations, the chemical potential was consistently updated. The energy cutoff \textit{encut} for all quantities was set to 5,000\,meV, except in the calculation for the energy shift and the $\mu$ update, where a reduced cutoff \textit{shiftcut} of 2,000\,meV was applied. 

In cases where both a DOS and a $W$ file are supplied, we renormalize $\mu$ based on the $\NF$ from the DOS. This not only ensures internal consistency but also facilitates improved convergence of $\mu$ with respect to the Brillouin zone grid sampling in our \textit{GW} computations.
When calculating \mustE{} from a provided $\mu$, the characteristic cutoff frequency for the phonon-introduced interaction $\omega_\text{ph}$ was set to the highest phonon frequency, while the characteristic electronic energy scale \textit{typEl} was defined by the width of the bands crossing the Fermi level.

\subsection*{Calculation of $W$}
Based on the DFT calculations in the normal state, the static polarizability function can be computed by applying the random phase approximation (RPA) as~\cite{PhysRevB.34.5390, DESLIPPE20121269} 
\begin{equation}
\label{eq:chiggp}
  \chi_{\mathbf{G}\mathbf{G}'}(\mathbf{q},0)=
  \frac{2}{N_{\mathrm{k}}}
  \sum_{n'}^{\text{occ.}}
  \sum_{n}^{\text{emp.}}
  \sum_{\mathbf{k}}
  \frac{\rho_{n\mathbf{k},n'\mathbf{k}+\mathbf{q}}(\mathbf{G})
  \left(\rho_{n\mathbf{k},n'\mathbf{k}+\mathbf{q}}(\mathbf{G}')\right)^*}{\varepsilon_{n'\mathbf{k}+\mathbf{q}}-\varepsilon_{n\mathbf{k}}}
\end{equation}
where $N_{\mathrm{k}}$ is the total number of wave vectors $\mathbf{k}$ in the first Brillouin zone, $\mathbf{G}$ is a reciprocal lattice vector, and
\begin{align}
  \rho_{n\mathbf{k},n'\mathbf{k}+\mathbf{q}}(\mathbf{G})
  =\int
  \psi^*_{n'\mathbf{k}+\mathbf{q}}(\mathbf{r})
  e^{i(\mathbf{q}+\mathbf{G})\cdot \mathbf{r}}
  \psi_{n\mathbf{k}}(\mathbf{r})
  d\mathbf{r}
\end{align} 
denotes the plane-wave matrix element. 
The inverse static dielectric function is expressed in terms of the static polarizability function $\chi_{\mathbf{G}\mathbf{G}'}$:
\begin{align}
\label{eq:epsggp}
  \epsilon^{-1}_{\mathbf{G}\mathbf{G}'}(\mathbf{q},0)
  =
  \delta_{\mathbf{G}\mathbf{G}'}
  -
  \frac{4\pi e^2}{\Omega}
  \frac{\chi_{\mathbf{G}\mathbf{G}'}(\mathbf{q},0)}{|\mathbf{q}+\mathbf{G}||\mathbf{q}+\mathbf{G}'|}.
\end{align} 
Since the screened Coulomb interaction is written with the Fourier transformation of $\epsilon^{-1}$ as
\begin{align}
 W(\mathbf{r}, \mathbf{r}', 0)=
 \int\!\mathrm{d}\mathbf{r}''
 \frac{\epsilon^{-1}(\mathbf{r}, \mathbf{r}'', 0)}{|\mathbf{r}''-\mathbf{r}'|},
\end{align} 
the screened Coulomb matrix element describing the scattering between two Kohn-Sham states is given by
\begin{align}
\label{eq:Wnknpkp}
  W_{n\mathbf{k},n'\mathbf{k}+\mathbf{q}}
  &=
  \int
  \int
  \psi^*_{n'\mathbf{k}+\mathbf{q}}(\mathbf{r})
  \psi_{n\mathbf{k}}(\mathbf{r})
  W(\mathbf{r}, \mathbf{r}', 0)
  \notag\\
  &\hspace{2em}\times\psi^*_{n'-\mathbf{k}-\mathbf{q}}(\mathbf{r}')
  \psi_{n-\mathbf{k}}(\mathbf{r}')
  d\mathbf{r} d\mathbf{r}'\notag\\
  &=
  \frac{4\pi e^2}{N_{\mathrm{q}}\Omega}
  \sum_{\mathbf{G},\mathbf{G}'}
  \epsilon^{-1}_{\mathbf{G}\mathbf{G}'}(\mathbf{q})
  \frac{
    \rho_{n\mathbf{k},n'\mathbf{k}+\mathbf{q}}(\mathbf{G})
    \left(\rho_{n\mathbf{k},n'\mathbf{k}+\mathbf{q}}(\mathbf{G}')\right)^*
  }{
    |\mathbf{q}+\mathbf{G}||\mathbf{q}+\mathbf{G}'|
  }.
\end{align}
Here, the relation $\psi_{n-\mathbf{k}}(\mathbf{r}) = \psi^*_{n\mathbf{k}}(\mathbf{r})$ is used, which holds if the Hamiltonian is invariant under time-reversal symmetry. 

In this work, we used BerkeleyGW~\cite{PhysRevB.34.5390,PhysRevB.62.4927,DESLIPPE20121269} to compute $\chi_{\mathbf{G}\mathbf{G}'}(\mathbf{q},0)$ and $\epsilon^{-1}_{\mathbf{G}\mathbf{G}'}(\mathbf{q},0)$ as defined in Eqs.~\eqref{eq:chiggp} and \eqref{eq:epsggp}, respectively. These quantities were evaluated for $\mathbf{G}$ and $\mathbf{G}'$ satisfying $|G^2| < |E_{\mathrm{cut}}|$, where $E_{\mathrm{cut}}$ is the dielectric energy cutoff and was set to 25\,Ry for all materials. An $8^3$ $\mathbf{q}$-grid was used for Nb, Al, and TiN; an $8\times8\times4$ $\mathbf{q}$-grid for Tc; and a $6^3$ $\mathbf{q}$-grid for NbC, H$_3$S, YH$_6$, LaBeH$_8$, and LaBH$_8$. The energy-dependent Coulomb interaction $W(\varepsilon,\varepsilon')$ was computed using Eqs.~\eqref{eq:aver_W} and \eqref{eq:Wnknpkp}.

\section*{Acknowledgments}
We thank P.N. Ferreira, F. Jöbstl, and M. Sahoo, for testing \IsoME{} and providing feedback. The authors acknowledge P.N. Ferreira for contributing data on H$_3$S and YH$_6$, and D. Khodachenko for data on LaBeH$_8$ and LaBH$_8$. We are also very grateful to A. Sanna for many fruitful discussions and sharing data for $W$ to compare and benchmark.
R.L. acknowledges the Carl Tryggers Stiftelse för Vetenskaplig Forskning (CTS 23: 2934). H.M. and E.R.M acknowledge the support from the U.S. National Science Foundation under Grant No. OAC-2103991. Z.Z. and Z.L. acknowledge the support from the U.S. National Science Foundation under Grant No. DMR-2440763.
Superconductivity computations were performed on local workstations and the lCluster at TU Graz. Calculations of the Coulomb interaction were performed on the Frontera supercomputer at the Texas
Advanced Computing Center via the Leadership Resource
Allocation No. DMR22004 and DMR22042. 

\section*{Author contributions}
E.K. and D.S. contributed equally. E.K., D.S., and C.H. developed the code and performed the superconductivity calculations, while R.L. derived the refined $\cp$ update procedure. H.M. and Z.Z. carried out the \textit{GW} calculations of the Coulomb interaction. C.H. conceived and supervised the project, with E.R.M. and Z.L. providing additional oversight. All authors contributed to discussions and revised the manuscript.

\section*{Declaration of competing interest}
The authors declare that they have no known competing financial interests or personal relationships that could have appeared to influence the
work reported in this paper.

\section*{Code Availability}
\IsoME{} is available as a registered Julia package (\href{https://juliahub.com/ui/Packages/General/IsoME}{juliahub.com/ui/Packages/General/IsoME}), on GitHub (\href{https://github.com/cheil/IsoME.jl}{github.com/cheil/IsoME.jl}), and Zenodo (\href{https://zenodo.org/records/14967551}{DOI:10.5281/zenodo.14967551}).

\section*{Data Availability}
Data is made available with the code. Any additional data will be made available upon request.

\section*{Declaration of generative AI and AI-assisted technologies in the writing process}

In preparing this paper, the authors used ChatGPT to help improve readability and language. After using this tool, the authors reviewed and edited the content as needed and take full responsibility for the content of the published article.

\FloatBarrier
\bibliographystyle{apsrev4-2}
\bibliography{literature}

\begin{thebibliography}{80}
\bibitem [{not({\natexlab{a}})}]{note_isoME}%
  \BibitemOpen
  \href@noop {} {} ({\natexlab{a}}),\ \bibinfo {note} {the IsoME code can be obtained via github at [...] or directly as a Julia package.}\BibitemShut {Stop}%
\end{thebibliography}

\clearpage
\newpage
\begin{figure*}
\includegraphics[page = 1, width=\linewidth]{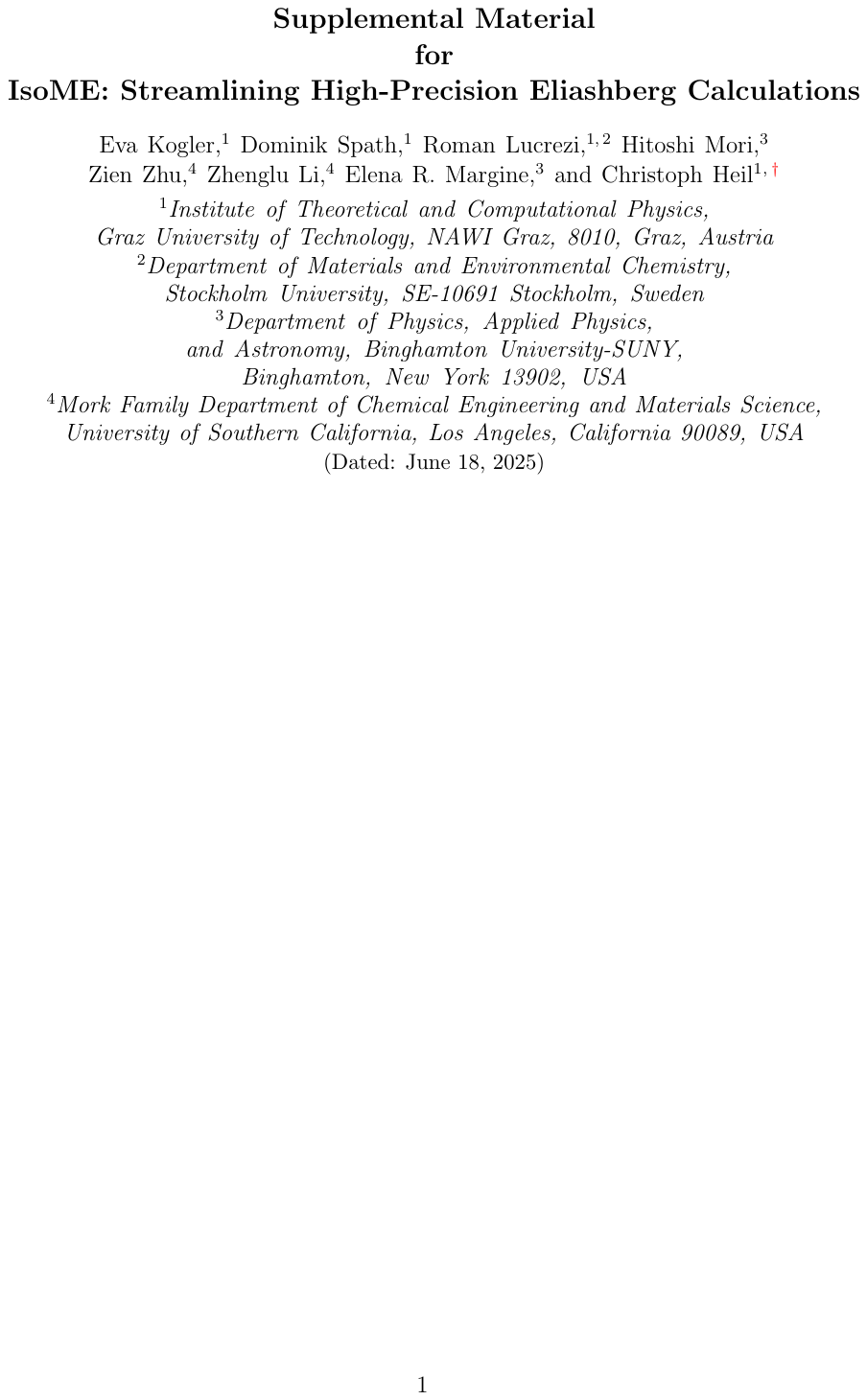}
\end{figure*}
\begin{figure*}
\includegraphics[page = 2, width=\linewidth]{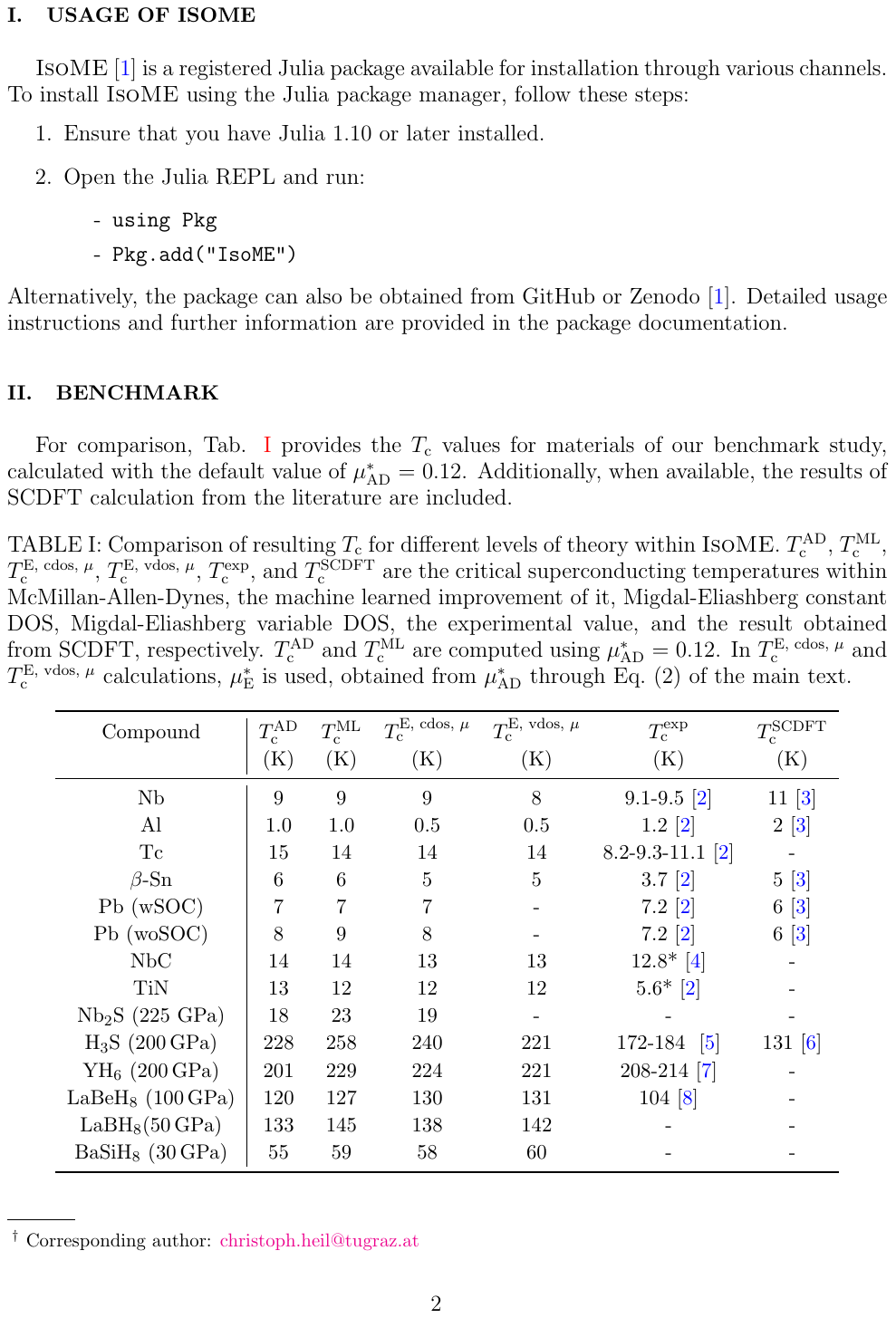}
\end{figure*}
\begin{figure*}
\includegraphics[page = 3, width=\linewidth]{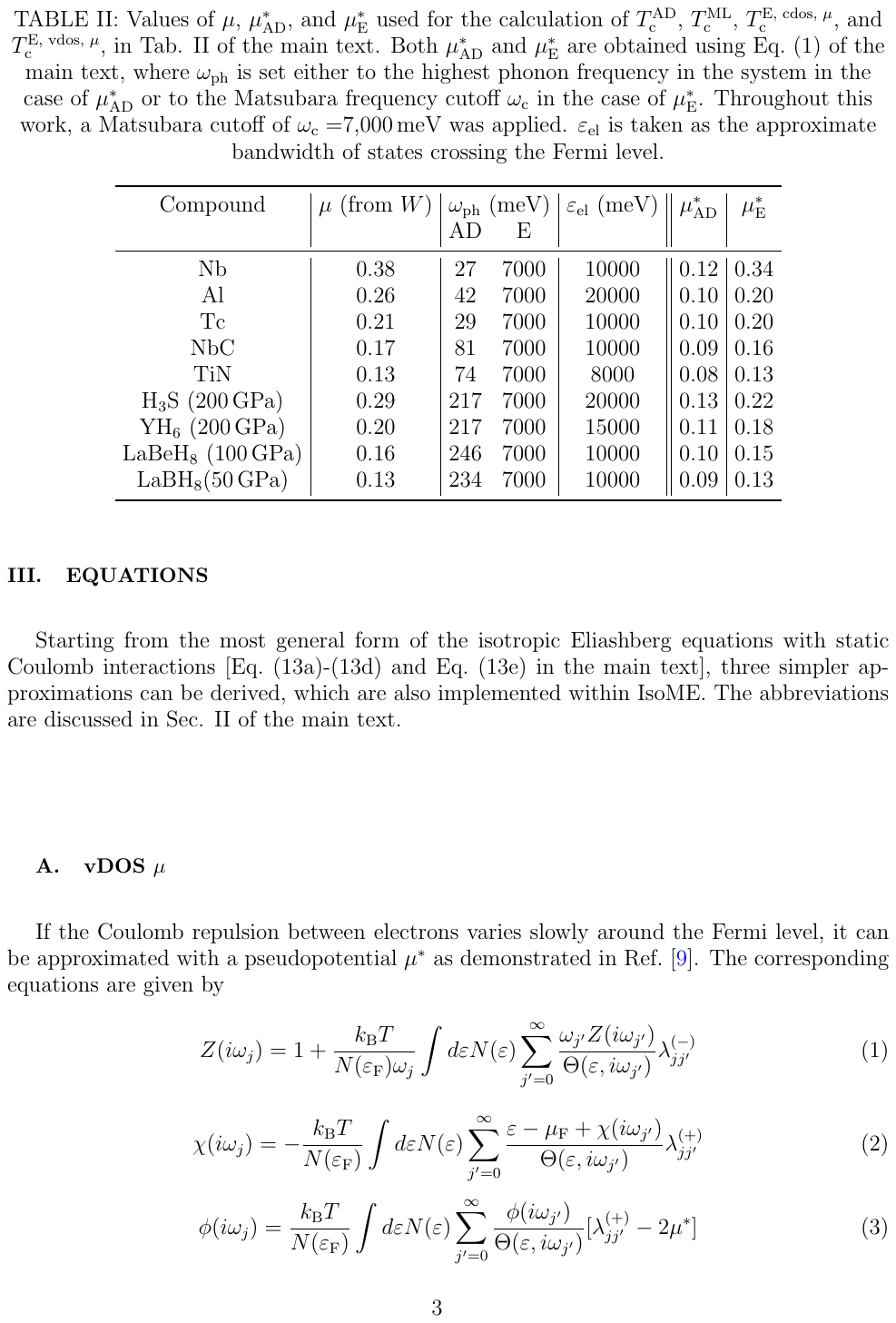}
\end{figure*}
\begin{figure*}
\includegraphics[page = 4, width=\linewidth]{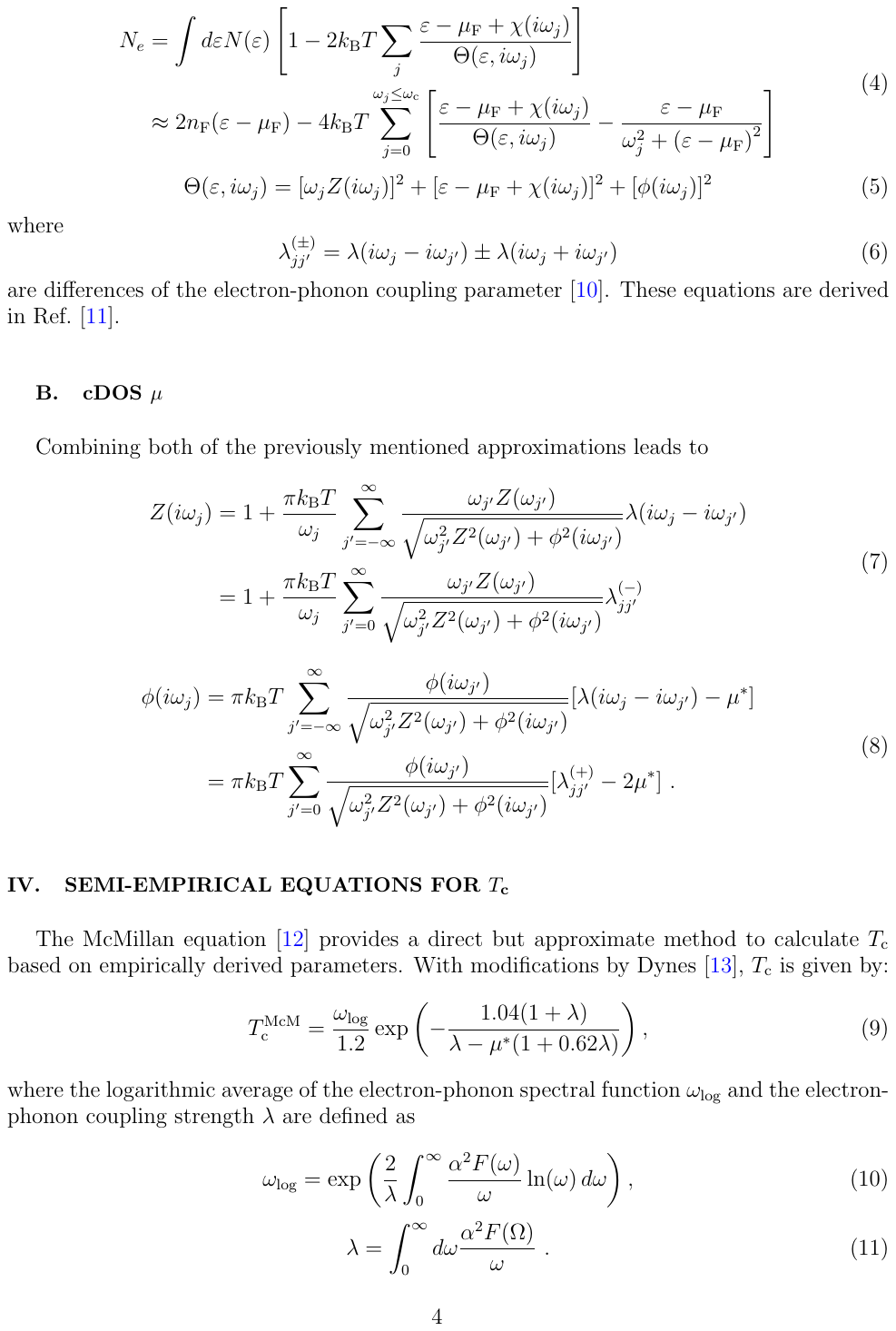}
\end{figure*}
\begin{figure*}
\includegraphics[page = 5, width=\linewidth]{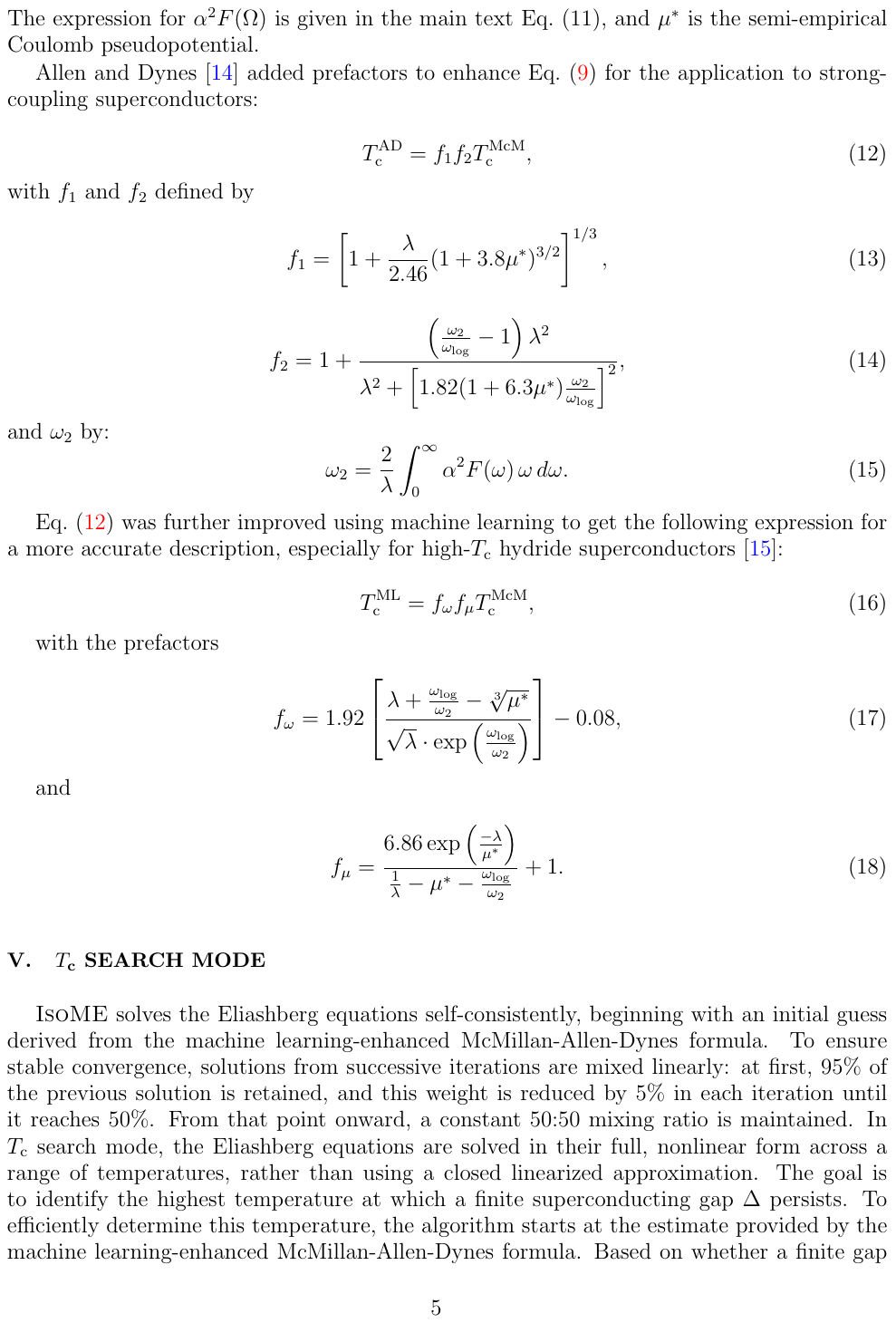}
\end{figure*}
\begin{figure*}
\includegraphics[page = 6, width=\linewidth]{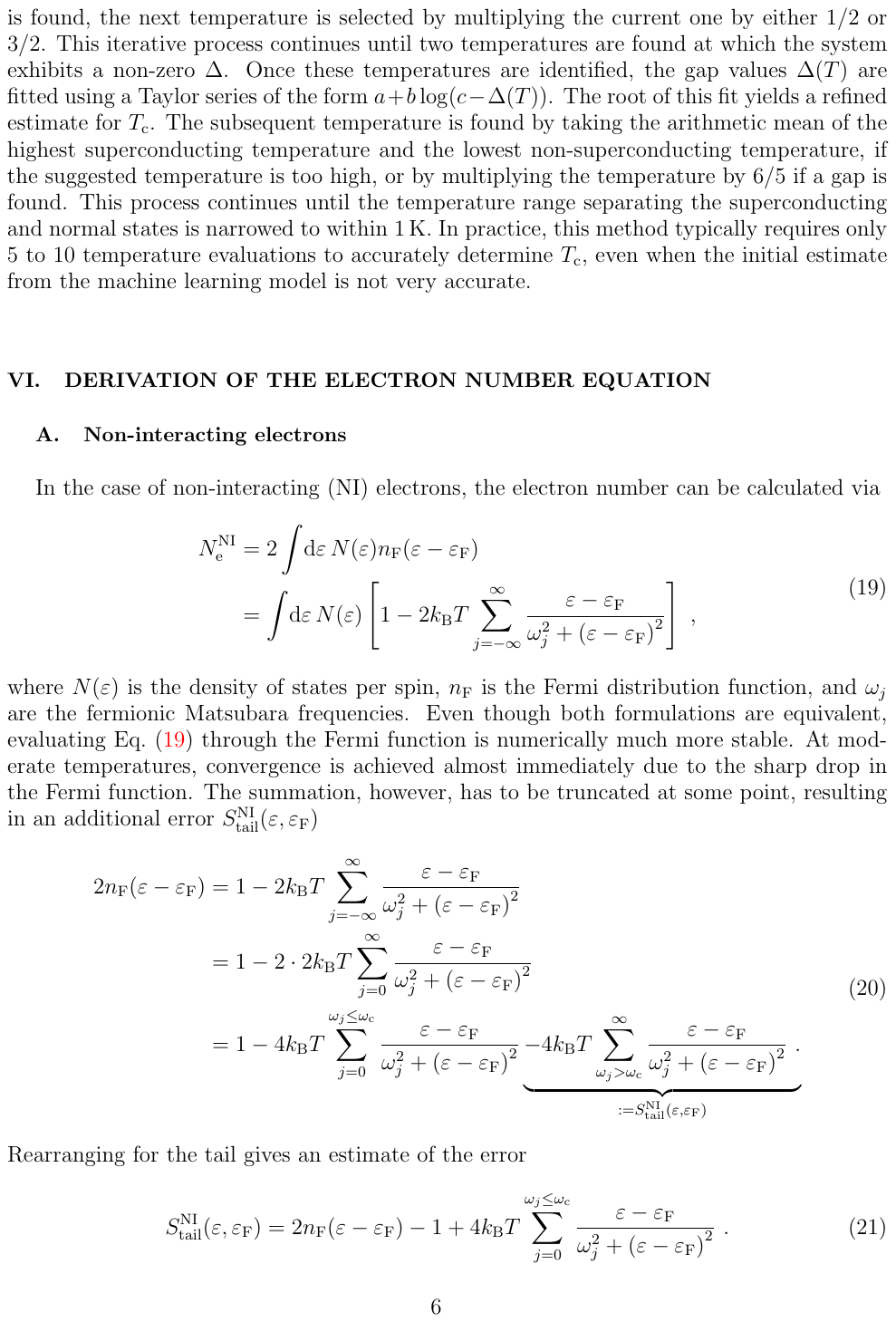}
\end{figure*}
\begin{figure*}
\includegraphics[page = 7, width=\linewidth]{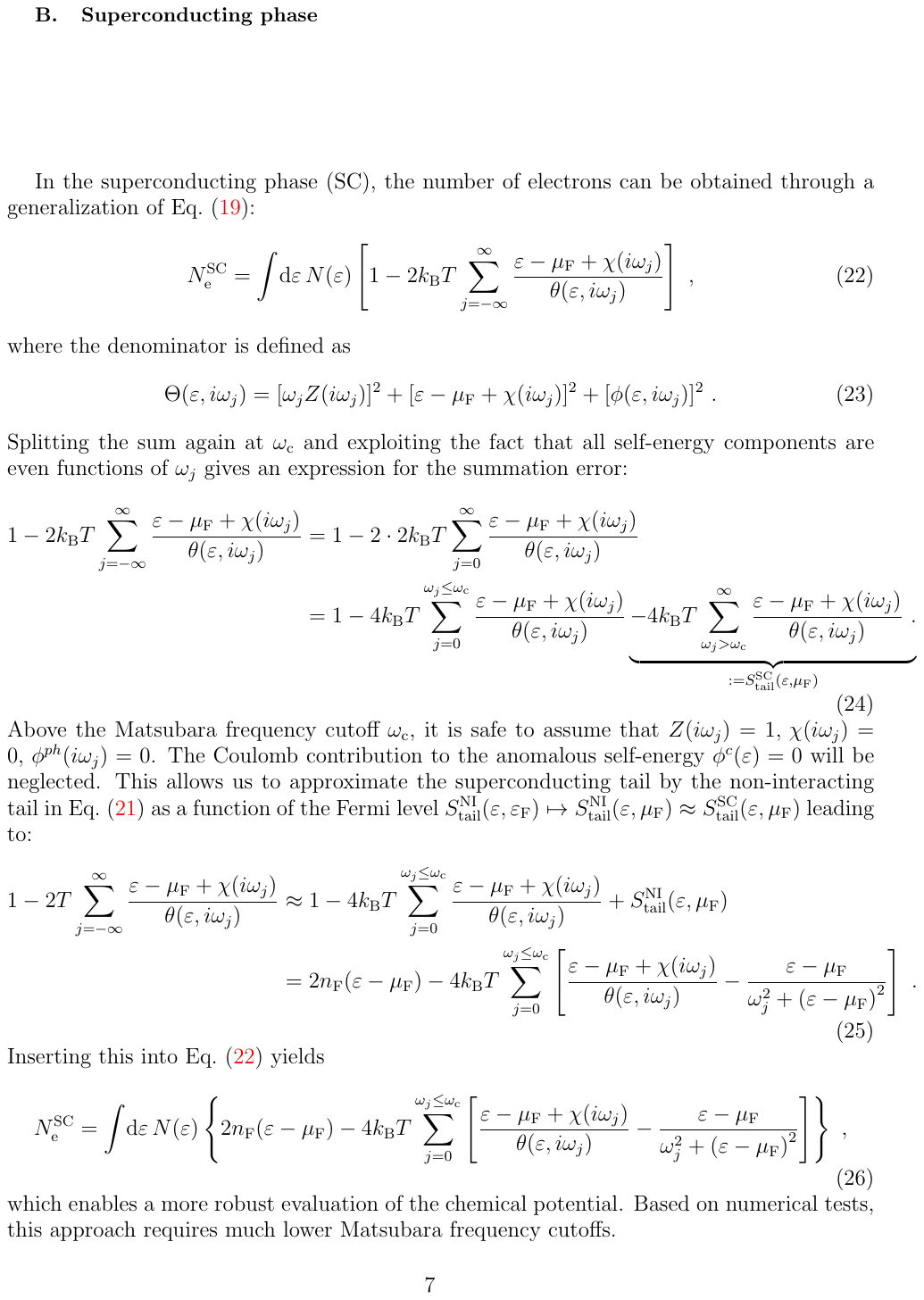}
\end{figure*}
\begin{figure*}
\includegraphics[page = 8, width=\linewidth]{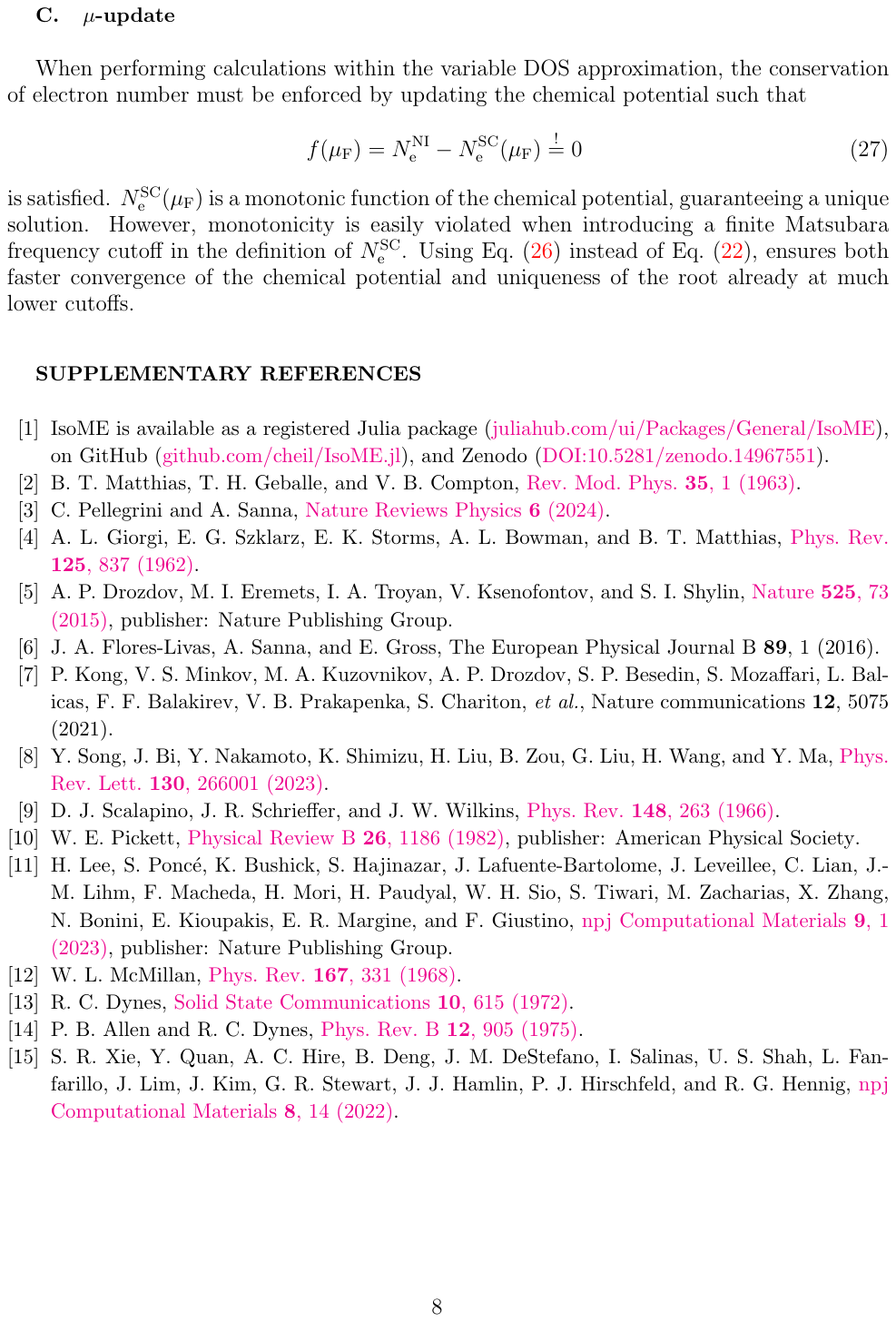}
\end{figure*}

\end{document}